\documentclass[preprint,11pt]{aastex}

\providecommand{\eg}{e.g.}
\providecommand{\etal}{et~al.}

\revised{}
\accepted{}

\slugcomment{\aj, in press}

\shorttitle{XRF~050416A}
\shortauthors{S.~T. Holland, et~al.}

\received{2006 August 14}
\begin{document}


\title{Optical, Infrared, and Ultraviolet Observations of the $X$-Ray
       Flash XRF~050416A}

\author{S.~T.~Holland\altaffilmark{1,2}
        P.~T.~Boyd\altaffilmark{3},
	J.~Gorosabel\altaffilmark{4},
	J.~Hjorth\altaffilmark{5},
        P.~Schady\altaffilmark{6,7},
	B.~Thomsen\altaffilmark{8},
        T.~Augusteijn\altaffilmark{9},
	A.~J.~Blustin\altaffilmark{6},
        A.~Breeveld\altaffilmark{6},
        M.~De~Pasquale\altaffilmark{6},
	J.~P.~U.~Fynbo\altaffilmark{5},
        N.~Gehrels\altaffilmark{3},
        C.~Gronwall\altaffilmark{7},
        S.~Hunsberger\altaffilmark{7},
        M.~Ivanushkina\altaffilmark{7,10},
        W.~Landsman\altaffilmark{3,11},
        P.~Laursen\altaffilmark{5},
        K.~McGowan\altaffilmark{6,12},
        V.~Mangano\altaffilmark{13},
	C.~B.~Markwardt\altaffilmark{3},
	F.~Marshall\altaffilmark{3},
        K.~O.~Mason\altaffilmark{6,14},
	A.~Moretti\altaffilmark{15},
        M.~J.~Page\altaffilmark{6},
        T.~Poole\altaffilmark{6},
        P.~Roming\altaffilmark{7},
        S.~Rosen\altaffilmark{6}, \&
        M.~Still\altaffilmark{2,3,16}}

\altaffiltext{1}{Code 660.1,
                 NASA's Goddard Space Flight Centre,
                 Greenbelt, MD 20771,
                 U.S.A.
                 \email{sholland@milkyway.gsfc.nasa.gov}}

\altaffiltext{2}{Universities Space Research Association}

\altaffiltext{3}{NASA's Goddard Space Flight Centre,
                 Greenbelt, MD 20771,
                 U.S.A.}

\altaffiltext{4}{Instituto de Astrof{\'\i}sica de Andaluc{\'\i}a (CSIC),
                 Apartado de Correos, 3004,
                 18080 Granada,
                 Spain}

\altaffiltext{5}{Dark Cosmology Centre,
                 Niels Bohr Institute,
                 University of Copenhagen,
                 Juliane Maries Vej 30,
                 DK--2100 Copenhagen \O,
                 Denmark}

\altaffiltext{6}{Mullard Space Science Laboratory,
                 Holmbury St Mary, Dorking, Surrey RH5~6NT, UK}

\altaffiltext{7}{Department of Astronomy and Astrophysics,
                 Pennsylvania State University,
                 525 Davey Laboratory,
                 University Park, PA 16802}

\altaffiltext{8}{Department of Physics and Astronomy,
                 University of Aarhus,
                 Ny Munkegade,
                 DK--8000 {\AA}rhus C,
                 Denmark}

\altaffiltext{9}{Nordic Optical Telescope,
                 Apartado 474,
                 38700 Santa Cruz de La Palma,
                 Canary Islands,
                 Spain}

\altaffiltext{10}{Department of Physics and Astronomy,
                  Brigham Young University, N208 ESC,
                  Provo, UT 84602,
                  USA}

\altaffiltext{11}{Science Systems \& Applications, Inc.}

\altaffiltext{12}{School of Physics \& Astronomy,
                  University of Southampton,
                  Southampton, Hampshire SO17~1BJ, UK}

\altaffiltext{13}{INAF--Istituto di Astrofisica Spaziale e Fisica Cosmica Sezione di Palermo,
                  via Ugo La Malfa 153
                  I--90146 Palermo,
                  Italy}

\altaffiltext{14}{PPARC, Polaris House,
                  North Star Ave,
                  Swindon SN2~1SZ, UK}

\altaffiltext{15}{INAF--Osservatorio Astronomico di Brera,
                  via Emilio Bianchi 46,
                  I--23807 Merate (Lc),
                  Italy}

\altaffiltext{16}{South African Astronomical Observatory,
                  PO Box 9, Observatory 7935
                  South Africa}


\begin{abstract}

     We present ultraviolet, optical, and infrared photometry of the
afterglow of the $X$-ray flash XRF~050416A taken between approximately
100 seconds and 36 days after the burst.  We find an intrinsic
spectral slope between 1930 {\AA} and 22\,200 {\AA} of $\beta = -1.14
\pm 0.20$ and a decay rate of $\alpha = -0.86 \pm 0.15$.  There is no
evidence for a change in the decay rate between approximately 0.7 and
4.7 days after the burst.  Our data implies that there is no spectral
break between the optical and $X$-ray bands between 0.7 and 4.7 days
after the burst, and is consistent with the cooling break being
redward of the $K_s$ band (22\,200 {\AA}) at 0.7 days.  The combined
ultraviolet/optical/infrared spectral energy distribution shows no
evidence for a significant amount of extinction in the host galaxy
along the line of sight to XRF~050416A.  Our data suggest that the
extragalactic extinction along the line of sight to the burst is only
approximately $A_V = 0.2$ mag, which is significantly less than the
extinction expected from the hydrogen column density inferred from
$X$-ray observations of XRF~050416A assuming a dust-to-gas ratio
similar to what is found for the Milky Way.  The observed extinction,
however, is consistent with the dust-to-gas ratio seen in the Small
Magellanic Cloud.  We suggest that XRF~050416A may have a
two-component jet similar to what has been proposed for GRB~030329.
If this is the case the lack of an observed jet break between 0.7 and
42 days is an illusion due to emission from the wide jet dominating
the afterglow after approximately 1.5 days.
\end{abstract}


\keywords{gamma rays: bursts}


\section{Introduction\label{SECTION:intro}}

     The {\sl Swift\/} satellite \citep{GCG2004} is a multi-instrument
observatory designed to detect and rapidly localize gamma-ray bursts
(GRBs).  The observatory contains three instruments.  The Burst Alert
Telescope (BAT\@; \citealt{Ba2005}) is used to identify GRBs and
localize them to $\la 3\arcmin$ in the energy range 15--150 keV.  Once
a burst has been detected {\sl Swift\/} slews to point its two
narrow-field instruments, the $X$-Ray Telescope (XRT\@;
\citealt{Bu2005}), and the UltraViolet/Optical Telescope (UVOT\@;
\citealt{R2005}) at the burst.  The XRT obtains $X$-ray localizations
to $\la 5\arcsec$ in the energy range 0.2--10 keV.  The UVOT
simultaneously obtains localizations to $\approx 0\farcs5$ then cycles
through a set of optical and ultraviolet filters.

     The $X$-ray flash (XRF) XRF~050416A was detected in the
constellation Coma Berenices by the BAT at 11:04:44.5 UTC
\citep{SBB2005a}.  The gamma-ray light curve showed a slowly rising
peak followed by several smaller peaks.  The burst had a $T_{90}$
duration of only $2.4 \pm 0.2$ s in the 15--150 keV band.
\citet{Sa2005} find a peak energy of $E_p = 15.6^{+2.3}_{-2.7}$ keV,
making XRF~050416A a soft burst, and they classify XRF~050416A as an
XRF.  The fluence was $(3.5 \pm 0.3) \times 10^{-7}$ erg cm$^{-2}$ in
the 15--150 keV band \citep{Sa2005}.  The fluence in the soft 15--25
keV band ($(1.7 \pm 0.2) \times 10^{-7}$ erg cm$^{-2}$) was larger
than that in the harder 50--100 keV band ($3.4^{+1.0}_{-0.6} \times
10^{-8}$ erg cm$^{-2}$).  This, and the low peak energy, makes
XRF~050416A an XRF.  A detailed analysis of the gamma-ray properties
of XRF~050416A is presented in \citet{Sa2005}.

     An $X$-ray afterglow was identified by \citet{KRB2005} and is
discussed in detail in \citet{MPC2006}.  The $X$-ray light curve
follows the canonical $X$-ray light curve described by \citet{NKG2006}
with an early-time decay of $\alpha_X = -2.4 \pm 0.5$ out to 172 s,
where the flux density, $f_\nu$, is related to the time since the BAT
trigger by $f_\nu(t) \propto t^\alpha$.  The canonical light curve has
an initially steep decay due to curvature effects in the relativistic
fireball.  This is followed by a slow decay due to energy injection.
The light curve makes a second transition to a decay slope of
approximately $-1$ when energy injection ends.  The slow decay section
for XRF~050416A has a slope of $\alpha_X = -0.44 \pm 0.13$ and lasts
from 172 to 1450 s.  The late-time decay has a slope of $\alpha_X =
-0.88 \pm 0.02$, which continues until at least 42 days after the BAT
trigger.  The $X$-ray spectrum after 1450 s has a slope of $\beta_X =
-1.04^{+0.05}_{-0.11}$, where the relationship between flux density
and frequency is $f_\nu \propto \nu^\beta$, and shows evidence for a
hydrogen column density in the host galaxy of $N_H = 6.8^{+1.0}_{-1.2}
\times 10^{21}$ cm$^{-2}$.  If we assume the \citet{PS1995}
relationship between hydrogen column density and extinction in the
Milky Way, $N_\mathrm{H} = (1.79 \times 10^{21}) A_V$, then this
column density implies $A_V = 3.8^{+0.6}_{-0.7}$ in the host galaxy
along the line of sight to XRF~050416A.

     The optical afterglow was first identified by \citet{CF2005}
using the Palomar 200-inch Hale Telescope and confirmed by
\citet{ASR2005} using the ANU 2.3-m telescope.  \citet{F2005} noted
that the afterglow is visible in the UVOT's UVW2 filter and stated
that this indicates a redshift of $z \la 1$.  A spectrum of the host
galaxy was obtained 51 days after the burst by \citet{CKG2005}.  They
measured a redshift for the host of $z = 0.6535 \pm 0.0002$ from
[OII], H$\beta$, H$\gamma$, and H$\delta$ emission lines.  The host's
blue colour suggests ongoing star formation, as is seen in many GRB
host galaxies.  UVOT magnitudes, based on preliminary photometric
calibrations, were published by \citet{SSM2005a,SSM2005b}.  This paper
supersedes those results.  \citet{S2005} found a 4.86 GHz radio
afterglow with a flux density of $260 \pm 55$ $\mu$Jy approximately
5.6 days after the BAT trigger.

     Excluding the short--hard bursts \citep{KMF1993} XRF~050416A,
with a redshift of 0.6535, is one of the nearest of the {\sl
Swift\/}-detected GRBs with spectral redshift determinations.  Its
$T_{90}$ duration and $E_p$ are similar to those of the XRF~050406
\citep{KBB2005,SMO2006} and are fairly typical of XRFs in general.
\citet{Sa2005} have shown that XRF~050416A is consistent with the
relationship between the isotropic energy equivalent,
$E_\mathrm{iso}$, and the peak energy in the rest frame,
$E_p^\prime$\footnote{Throughout this paper we use primes to indicate
quantities in the rest frame.}, found by \citet{AFT2002} for classical
long--soft GRBs.  This suggests that XRFs and GRBs are intimately
related.  XRF~050416A does not, however, follow the the empirical
relation between $E_p^\prime$, $E_\mathrm{iso}$, and the jet break
time in the rest frame, $t_j^\prime$ of \citet{LZ2005}, which suggests
that either this XRF has an unusually late jet break or that something
is preventing us from seeing the jet break, such as a long-lasting
emission component.  Something similar was seen in GRB~021004, where
identifying the jet break was complicated by several emission
components \citep{HWF2003}.

     In this paper we present space- and ground-based ultraviolet,
optical, and infrared observations of XRF~050416A.  We have adopted a
cosmology with a Hubble parameter of $H_0 = 70$
km~s$^{-1}$~Mpc$^{-1}$, a matter density of $\Omega_m = 0.3$, and a
cosmological constant of $\Omega_\Lambda = 0.7$.  For this cosmology a
redshift of $z = 0.6535$ corresponds to a luminosity distance of 3917
Mpc and a distance modulus of 42.96.  One arcsecond corresponds to
11.48 comoving kpc, or 6.95 proper kpc.  The look back time is 6.05
Gyr.


\section{Observations\label{SECTION:obs}}

     The optical afterglow for XRF~050416A is located at R.A. =
12:33:54.6, Dec.\ = +21:03:27 (J2000) \citep{CF2005}, which
corresponds to Galactic coordinates of $(b^\mathrm{II},l^\mathrm{II})$
= $(+82\fdg7316,268\fdg7316)$.  The reddening maps of \citet{SFD1998}
give a Galactic reddening of $E_{B-V} = 0.03 \pm 0.02$~mag in this
direction.  The corresponding Galactic extinctions are $A_U = 0.16$,
$A_B = 0.13$, $A_V = 0.10$, $A_{R_C} = 0.08$, $A_{I_C} = 0.06$, and
$A_{K_s} = 0.01$.  The Galactic extinctions in the UVOT ultraviolet
filters were calculated using the Milky Way extinction law from
\citet{P1992}.  The ultraviolet extinctions are $A_\mathrm{UVW1} =
0.23$, $A_\mathrm{UVM2} = 0.29$, and $A_\mathrm{UVW2} = 0.21$.

\begin{figure}
\plotone{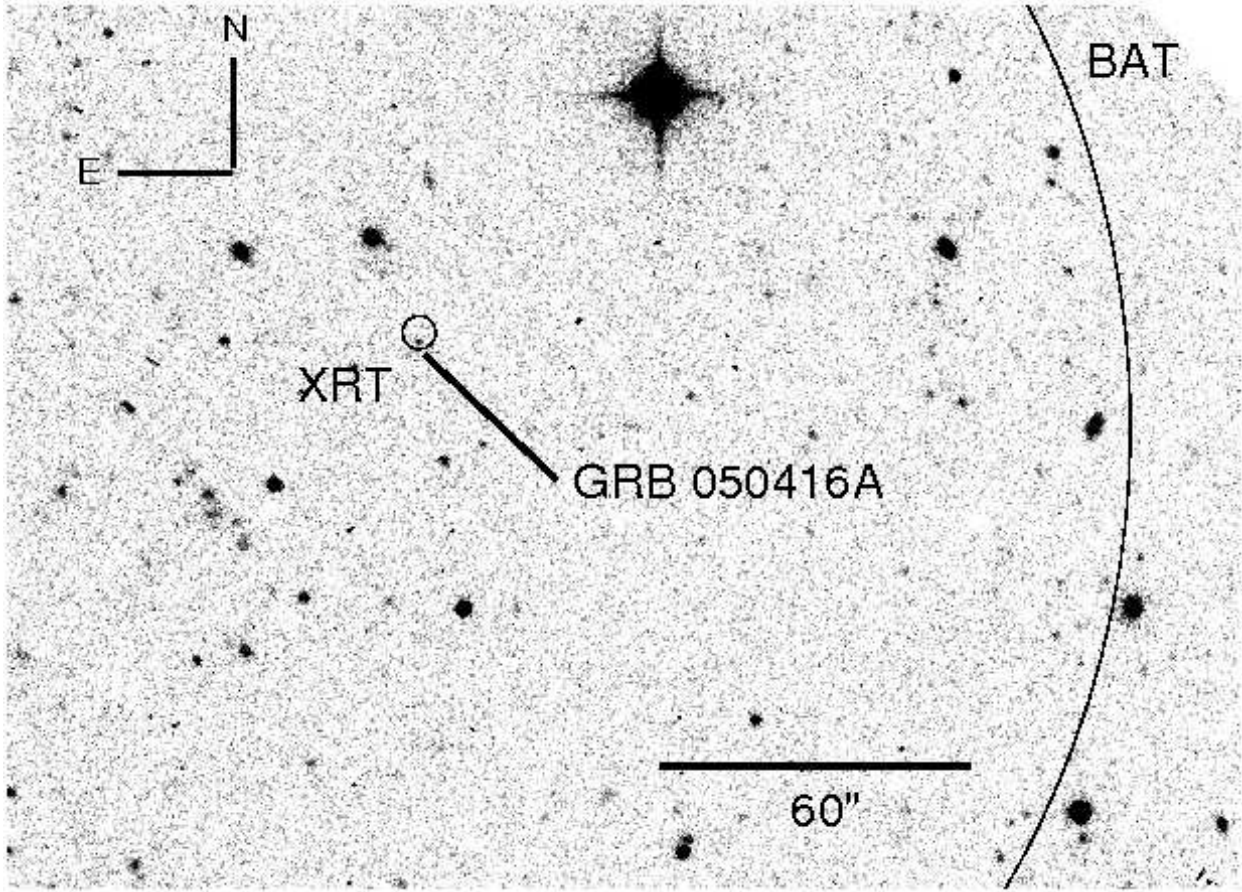} \figcaption[./f1.eps]{This figure shows the $2
\times 1200$ second DFOSC $V$-band image taken 60\,415 seconds after
the BAT trigger.  The large circle is the 3{\arcmin} radius BAT error
circle \citep{SBB2005b}.  The small circle is the $3\farcs3$ radius
XRT error circle \citep{MPC2006}.  North is up and east is to the
left.\label{FIGURE:field}}
\end{figure}

     The field of XRF~050416A is shown in Figure~\ref{FIGURE:field}.

\subsection{UVOT Data\label{SECTION:uvot}}

     The {\sl Swift\/} spacecraft slewed promptly when the BAT
detected XRF~050416A, and UVOT began imaging the field 65 s after the
BAT trigger. A range of 14--16 exposures were taken in each UVOT
filter between 11:05:49 UTC on 16 Apr 2005 and 03:45:46 UTC on 18 Apr
2005.  The UVOT has $U\!BV$ filters which approximate the Johnson
system, and three ultraviolet filters: UVW1 with a central wavelength
of $\lambda_0 = 2600$ \AA, UVM2 with $\lambda_0 = 2200$ \AA, and UVW2
with $\lambda_0 = 1930$ {\AA}.  Examination of the first settled
observation (a 100 s full-frame exposure in $V$) revealed a new source
relative to the Digital Sky Survey inside the XRT error circle.  This
source had a magnitude of $V = 19.19 \pm 0.29$.  The UVOT position of
the source is R.A.\ = 12:33:54.596, Dec.\ = +21:03:26.07 (J2000) with
a statistical error of $0\farcs01$ and an absolute astrometric
accuracy of $0\farcs56$ (90\% containment).  This position is
$0\farcs3$ from the reported XRT position \citep{MPC2006}.  It is
inside the XRT error circle and consistent with the ground-based
detection reported by \citep{CF2005}. Subsequent exposures showed the
source to be rapidly fading.

\begin{figure}
\plotone{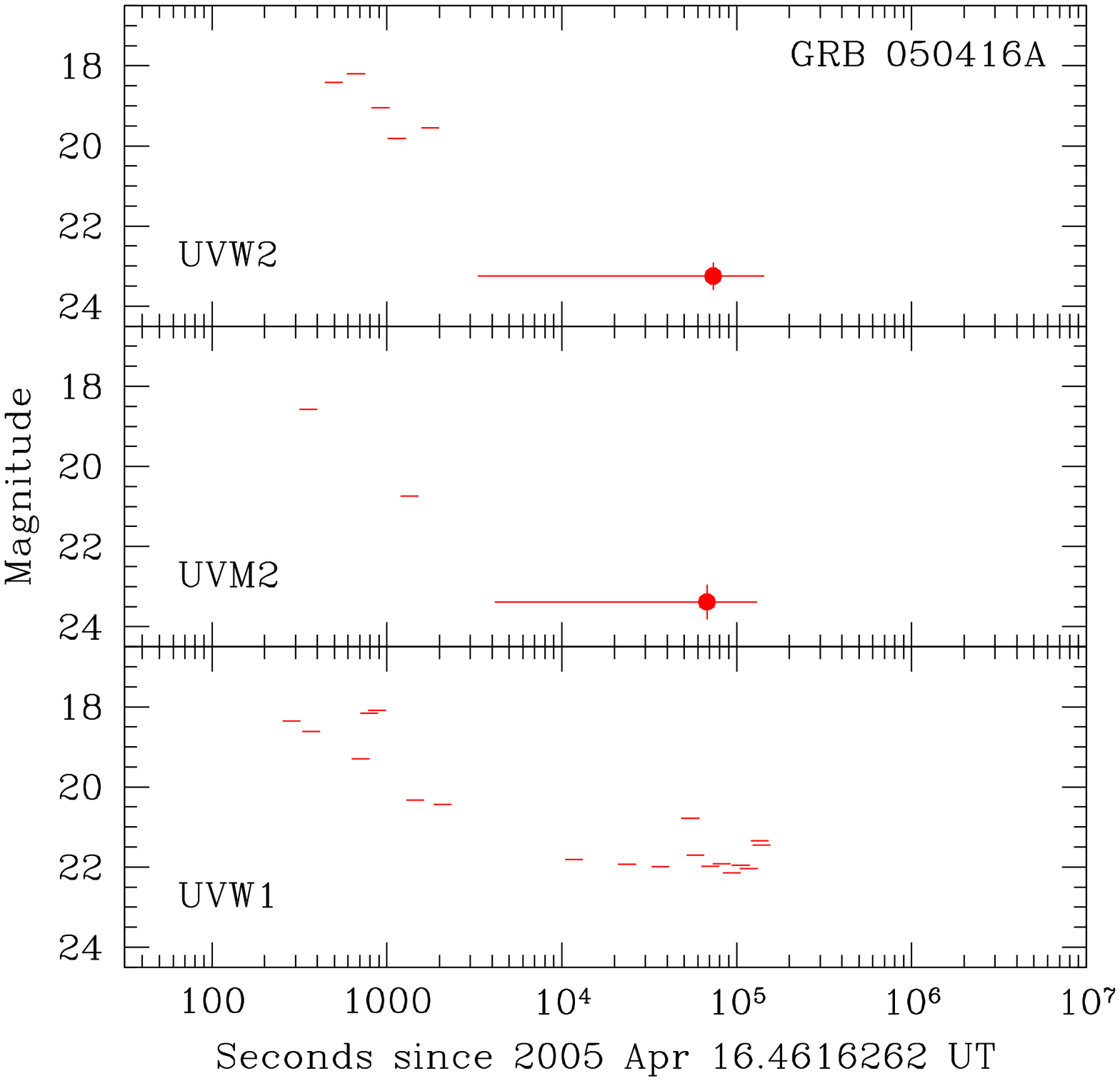} \figcaption[./f2.eps]{This Figure shows the UVW2,
UVM2, and UVW1 photometry of the afterglow of XRF~050416A.  Red
circles indicates UVOT detections.  Upper limits are represented by
horizontal lines.  The horizontal error bars on the detections
indicate the total period during which data was collected, not the
exposure time.\label{FIGURE:data1}}
\end{figure}

\begin{figure}
\plotone{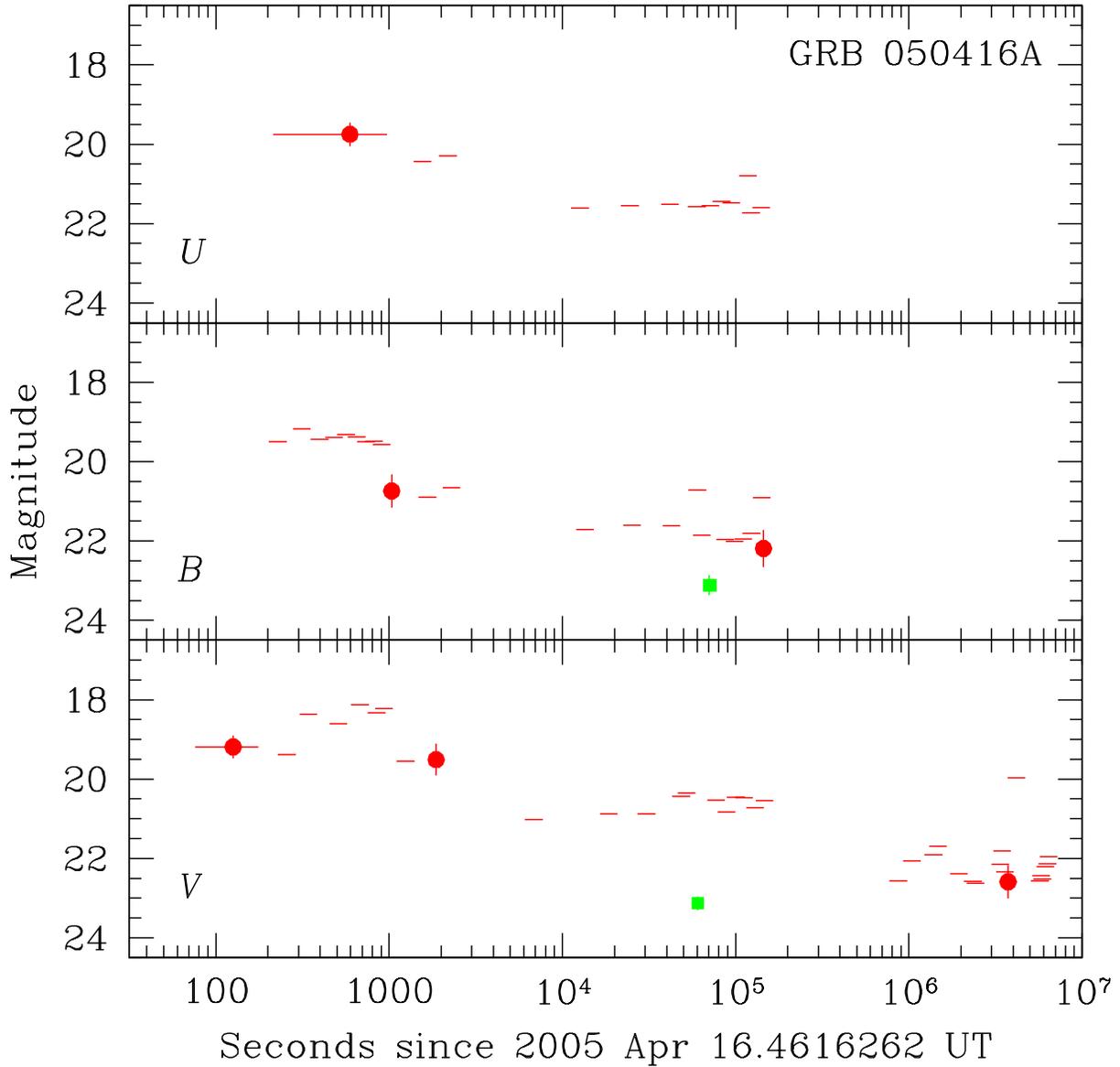} \figcaption[./f3.eps]{This Figure shows the UVOT
(red, circles) and DFOSC (green, squares) $U$, $B$, and $V$ photometry
of the afterglow of XRF~050416A.  The details are the same as for
Figure~\ref{FIGURE:data1}.\label{FIGURE:data2}}
\end{figure}

\begin{figure}
\plotone{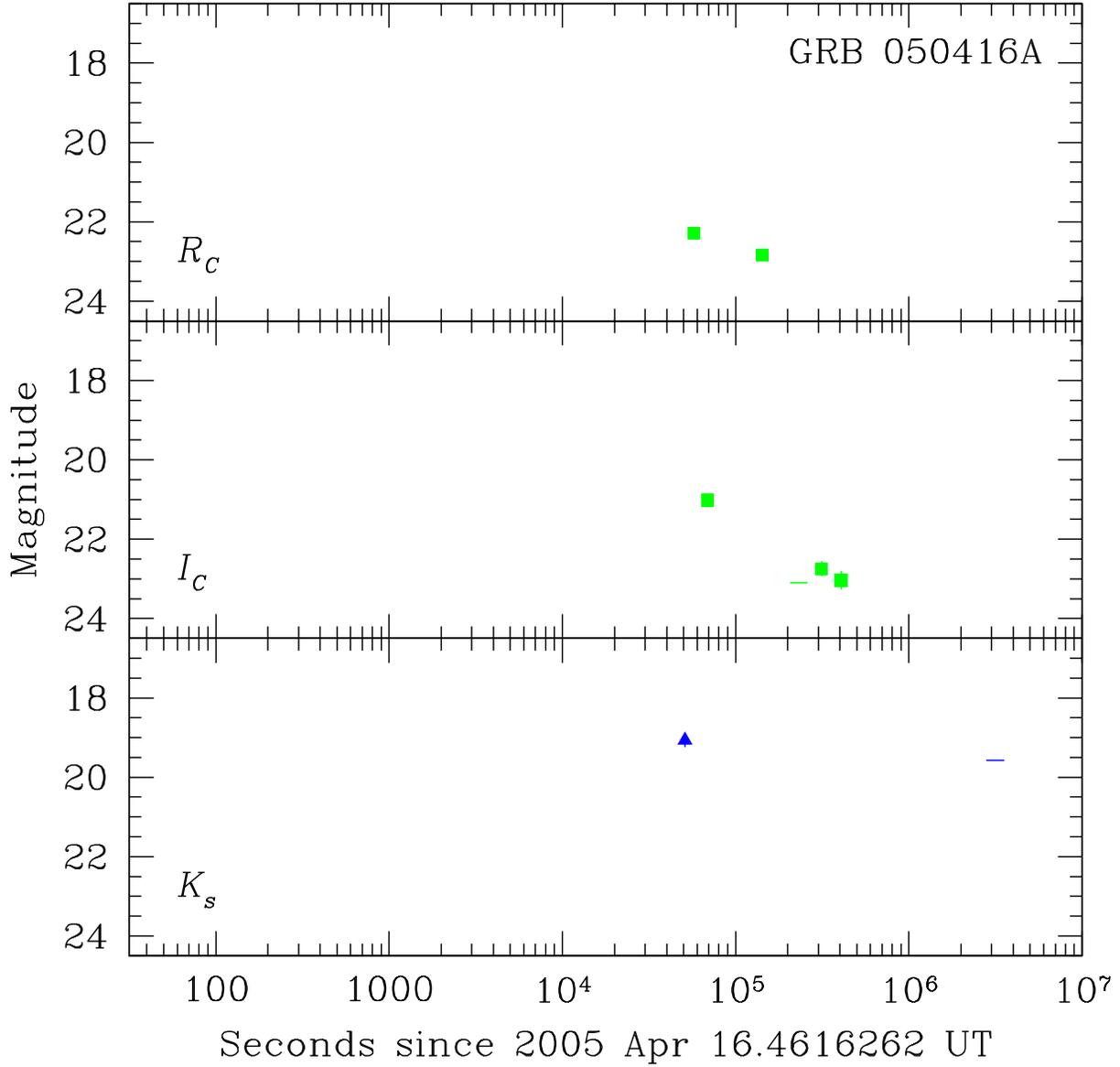} \figcaption[./f4.eps]{This Figure shows the DFOSC
(green, squares) and NOTCam (blue, triangles) $R_C$, $I_C$, and $K_s$
photometry of the afterglow of XRF~050416A.  The details are the same
as for Figure~\ref{FIGURE:data1}.\label{FIGURE:data3}}
\end{figure}

     The UVOT took 342 exposures of the field containing XRF~050416A
between 2005 Apr 16 and 2005 May 13 UTC\@.  For the vast majority of
these exposures the afterglow was too faint to be detected.  The UVOT
photometry that was used in this paper is shown in
Figures~\ref{FIGURE:data1},~\ref{FIGURE:data2},~and~\ref{FIGURE:data3}
and presented in Table~\ref{TABLE:phot}.  Magnitudes have not been
corrected for extinction.  All upper limits reported in
Table~\ref{TABLE:phot} are 3$\sigma$ and were obtained using a
$2\arcsec$ radius aperture with aperture corrections applied as
discussed below.

\clearpage

\begin{deluxetable}{rrcrrl|rrcrrl}
\tabletypesize{\scriptsize}
\tablewidth{0pt}
\tablecaption{Photometry of XRF~050416A.  The time $t_\mathrm{mid}$ is
the time from the BAT trigger to the middle of the observation.  The
exposure time is denoted by $\Delta t$.  Upper limits are 3$\sigma$
upper limits.  Errors are 1$\sigma$ statistical errors and do not
include the systematic errors in the photometric zero
points.\label{TABLE:phot}}

\tablehead{%
        \colhead{$t_\mathrm{mid}$ (s)} &
        \colhead{$\Delta t$ (s)} &
        \colhead{Filter} &
        \colhead{Mag} &
        \colhead{Err} &
        \colhead{Instrument} &
        \colhead{$t_\mathrm{mid}$ (s)} &
        \colhead{$\Delta t$ (s)} &
        \colhead{Filter} &
        \colhead{Mag} &
        \colhead{Err} &
        \colhead{Instrument}}
\startdata
         126 &             100  &   $V$ &   19.19  & 0.29    & UVOT &    59\,659 &             900  &   $U$ & $>21.57$ & \nodata & UVOT \\
         228 &              10  &   $B$ & $>19.49$ & \nodata & UVOT &    60\,166 &              99  &   $B$ & $>20.72$ & \nodata & UVOT \\
         257 &              10  &   $V$ & $>19.39$ & \nodata & UVOT &    60\,415 &  $2 \times 1200$ &   $V$ &   23.13  & 0.17    & DFOSC\tablenotemark{a} \\
         285 &              10  &  UVW1 & $>18.36$ & \nodata & UVOT &    63\,859 &             900  &   $B$ & $>21.86$ & \nodata & UVOT \\
         313 &              10  &   $B$ & $>19.17$ & \nodata & UVOT &    67\,559 &            9202  &  UVM2 &   23.38  & 0.42    & UVOT\tablenotemark{a} \\
         342 &              10  &   $V$ & $>18.36$ & \nodata & UVOT &    68\,693 &  $3 \times  600$ & $I_C$ &   21.02  & 0.17    & DFOSC\tablenotemark{a} \\
         355 &              10  &  UVM2 & $>18.57$ & \nodata & UVOT &    70\,541 &             900  &  UVW1 & $>21.98$ & \nodata & UVOT \\
         369 &              10  &  UVW1 & $>18.61$ & \nodata & UVOT &    70\,549 &            1500  &   $B$ &   23.11  & 0.25    & DFOSC \\
         397 &              10  &   $B$ & $>19.44$ & \nodata & UVOT &    71\,393 &             791  &   $U$ & $>21.55$ & \nodata & UVOT \\
         482 &              10  &   $B$ & $>19.39$ & \nodata & UVOT &     73\,345 &         10\,146  &  UVW2 &   23.25  & 0.33    & UVOT\tablenotemark{a} \\
         497 &              10  &  UVW2 & $>18.41$ & \nodata & UVOT &     76\,988 &             459  &   $V$ & $>20.53$ & \nodata & UVOT \\
         510 &              10  &   $V$ & $>18.61$ & \nodata & UVOT &     82\,070 &             900  &  UVW1 & $>21.92$ & \nodata & UVOT \\
         566 &              10  &   $B$ & $>19.32$ & \nodata & UVOT &     82\,944 &             835  &   $U$ & $>21.44$ & \nodata & UVOT \\
         594 & $10 \times   10$ &   $U$ &   19.75  & 0.30    & UVOT\tablenotemark{a} &     86\,973 &             900  &   $B$ & $>21.96$ & \nodata & UVOT \\
         650 &              10  &   $B$ & $>19.38$ & \nodata & UVOT &     88\,743 &             810  &   $V$ & $>20.83$ & \nodata & UVOT \\
         666 &              10  &  UVW2 & $>18.20$ & \nodata & UVOT &     93\,637 &             900  &  UVW1 & $>22.15$ & \nodata & UVOT \\
         679 &              10  &   $V$ & $>18.13$ & \nodata & UVOT &     94\,514 &             841  &   $U$ & $>21.47$ & \nodata & UVOT \\
         707 &              10  &  UVW1 & $>19.30$ & \nodata & UVOT &     98\,641 &             900  &   $B$ & $>22.01$ & \nodata & UVOT \\
         735 &              10  &   $B$ & $>19.49$ & \nodata & UVOT &    100\,162 &             311  &   $V$ & $>20.45$ & \nodata & UVOT \\
         792 &              10  &  UVW1 & $>18.16$ & \nodata & UVOT &    105\,335 &             900  &  UVW1 & $>21.96$ & \nodata & UVOT \\
         819 &              10  &   $B$ & $>19.48$ & \nodata & UVOT &    110\,415 &             900  &   $B$ & $>21.95$ & \nodata & UVOT \\
         848 &              10  &   $V$ & $>18.33$ & \nodata & UVOT &    112\,038 &             514  &   $V$ & $>20.47$ & \nodata & UVOT \\
         876 &              10  &  UVW1 & $>18.09$ & \nodata & UVOT &    117\,400 &             900  &  UVW1 & $>22.04$ & \nodata & UVOT \\
         904 &              10  &   $B$ & $>19.57$ & \nodata & UVOT &    117\,970 &             223  &   $U$ & $>20.80$ & \nodata & UVOT \\
         919 &              10  &  UVW2 & $>19.05$ & \nodata & UVOT &    122\,543 &             900  &   $U$ & $>21.73$ & \nodata & UVOT \\
         933 &              10  &   $V$ & $>18.22$ & \nodata & UVOT &    123\,434 &             868  &   $B$ & $>21.80$ & \nodata & UVOT \\
        1035 &             100  &   $B$ &   20.74  & 0.42    & UVOT &    129\,343 &             624  &   $V$ & $>20.72$ & \nodata & UVOT \\
        1140 &             100  &  UVW2 & $>19.81$ & \nodata & UVOT &    135\,246 &             390  &  UVW1 & $>21.34$ & \nodata & UVOT \\
        1245 &             100  &   $V$ & $>19.55$ & \nodata & UVOT &    138\,805 &             401  &  UVW1 & $>21.45$ & \nodata & UVOT \\
        1349 &             100  &  UVM2 & $>20.75$ & \nodata & UVOT &    140\,587 &             900  &   $U$ & $>21.60$ & \nodata & UVOT \\
        1454 &             100  &  UVW1 & $>20.32$ & \nodata & UVOT &    141\,137 &             184  &   $B$ & $>20.91$ & \nodata & UVOT \\
        1558 &             100  &   $U$ & $>20.44$ & \nodata & UVOT &    142\,008 & $25 \times  400$ & $R_C$ &   22.84  & 0.11    & DFOSC\tablenotemark{a} \\
        1663 &             100  &   $B$ & $>20.90$ & \nodata & UVOT &    144\,815 &             900  &   $B$ &   22.19  & 0.47    & UVOT \\
        1768 &             100  &  UVW2 & $>19.55$ & \nodata & UVOT &    146\,738 &             553  &   $V$ & $>20.55$ & \nodata & UVOT \\
        1868 &              93  &   $V$ &   19.51  & 0.40    & UVOT &    231\,672 & $53 \times  300$ & $I_C$ & $>23.10$ & \nodata & DFOSC\tablenotemark{a} \\
        2081 &             100  &  UVW1 & $>20.43$ & \nodata & UVOT &    313\,726 & $29 \times  300$ & $I_C$ &   22.74  & 0.19    & DFOSC\tablenotemark{a} \\
        2185 &             100  &   $U$ & $>20.30$ & \nodata & UVOT &    406\,438 & $21 \times  200$ & $I_C$ &   23.03  & 0.23    & DFOSC\tablenotemark{a} \\
        2290 &             100  &   $B$ & $>20.65$ & \nodata & UVOT &    868\,505 &          20\,957 &   $V$ & $>22.56$ & \nodata & UVOT\tablenotemark{a} \\
        6850 &             900  &   $V$ & $>21.02$ & \nodata & UVOT &    1039340  &             8274 &   $V$ & $>22.06$ & \nodata & UVOT\tablenotemark{a} \\
     11\,730 &             900  &  UVW1 & $>21.81$ & \nodata & UVOT &    1384870  &             6755 &   $V$ & $>21.91$ & \nodata & UVOT\tablenotemark{a} \\
     12\,637 &             900  &   $U$ & $>21.61$ & \nodata & UVOT &    1468840  &             4722 &   $V$ & $>21.69$ & \nodata & UVOT\tablenotemark{a} \\
     13\,508 &             826  &   $B$ & $>21.71$ & \nodata & UVOT &    1941288  &          14\,167 &   $V$ & $>22.38$ & \nodata & UVOT\tablenotemark{a} \\
     18\,532 &             900  &   $V$ & $>20.88$ & \nodata & UVOT &    2338167  &          21\,133 &   $V$ & $>22.58$ & \nodata & UVOT\tablenotemark{a} \\
     23\,601 &             900  &  UVW1 & $>21.93$ & \nodata & UVOT &    2424904  &          26\,389 &   $V$ & $>22.63$ & \nodata & UVOT\tablenotemark{a} \\
     24\,508 &             900  &   $U$ & $>21.55$ & \nodata & UVOT & 3\,146\,746 & $39 \times   52$ & $K_s$ & $>19.57$ & \nodata & NOTCam\tablenotemark{a} \\ 
     25\,231 &             530  &   $B$ & $>21.60$ & \nodata & UVOT & 3\,371\,121 &          12\,866 &   $V$ & $>22.15$ & \nodata & UVOT\tablenotemark{a} \\
     30\,599 &             900  &   $V$ & $>20.88$ & \nodata & UVOT & 3\,455\,561 &             4722 &   $V$ & $>21.81$ & \nodata & UVOT\tablenotemark{a} \\
     36\,640 &             858  &  UVW1 & $>22.00$ & \nodata & UVOT & 3\,590\,733 &          15\,986 &   $V$ & $>22.34$ & \nodata & UVOT\tablenotemark{a} \\
     41\,791 &             900  &   $U$ & $>21.51$ & \nodata & UVOT & 3\,732\,868 &          18\,806 &   $V$ &   22.59  & 0.42    & UVOT\tablenotemark{a} \\
     42\,552 &             606  &   $B$ & $>21.61$ & \nodata & UVOT & 4\,170\,209 &              321 &   $V$ & $>19.97$ & \nodata & UVOT\tablenotemark{a} \\
     48\,456 &             373  &   $V$ & $>20.44$ & \nodata & UVOT & 5\,706\,808 &          25\,285 &   $V$ & $>22.57$ & \nodata & UVOT\tablenotemark{a} \\
     51\,023 & $36 \times   78$ & $K_s$ &   19.07  & 0.17    & NOTCam\tablenotemark{a} & 5\,793\,106 &          20\,898 &   $V$ & $>22.43$ & \nodata & UVOT\tablenotemark{a} \\
     52\,031 &             271  &   $V$ & $>20.35$ & \nodata & UVOT & 5\,880\,015 &          23\,533 &   $V$ & $>22.52$ & \nodata & UVOT\tablenotemark{a} \\
     54\,363 &             132  &  UVW1 & $>20.78$ & \nodata & UVOT & 6\,136\,666 &          15\,285 &   $V$ & $>22.21$ & \nodata & UVOT\tablenotemark{a} \\
     57\,086 &  $6 \times  900$ & $R_C$ &   22.29  & 0.08    & DFOSC\tablenotemark{a}  & 6\,310\,565 &          13\,933 &   $V$ & $>22.13$ & \nodata & UVOT\tablenotemark{a} \\
     57\,990 &             773  &  UVW1 & $>21.70$ & \nodata & UVOT & 6\,394\,649 &          10\,004 &   $V$ & $>21.95$ & \nodata & UVOT\tablenotemark{a} \\
\enddata
\tablenotetext{a}{This data point consists of coadded data.}
\end{deluxetable}

\clearpage

     We performed photometry on each UVOT exposure using a circular
aperture with a radius of $2\arcsec$ centred on the position of the
optical afterglow.  This radius is approximately equal to the
full-width at half-maximum (FWHM) of the UVOT point-spread function
(PSF).  The PSF varies with filter and with the temperature of the
telescope, so we did not match the extraction aperture to the PSF for
each exposure.  The PSF FWHM, averaged over the temperature
variations, ranges from $1\farcs79 \pm 0\farcs05$ for the $V$ filter
to $2\farcs17 \pm 0\farcs03$ for the UVW2 filter.  The background was
measured in a sky annulus of inner radius $17\farcs5$ and width
$5\arcsec$ centred on the afterglow.  The source was not detected in
the individual ultraviolet filter exposures, so these exposures were
coadded.  The source was detected in the coadded UVW2 and UVM2
exposures, but not in the coadded UVW1 exposure.

     Aperture corrections were computed for each exposure to convert
the $2\arcsec$ photometry to the standard aperture radii used to
define UVOT's photometric zero points ($6\arcsec$ for $U\!BV$ and
$12\arcsec$ for the ultraviolet filters).  Six isolated stars were
used to compute the aperture correction for each exposure.  The RMS
scatter in the mean aperture correction for a single exposure was
typically $\approx 0.02$ mag.  The RMS scatter for each exposure was
added in quadrature to the statistical error in the $2\arcsec$
magnitude to obtain the total 1$\sigma$ error in each point.  All
detections above the 2$\sigma$ significance level are tabulated in
Table~\ref{TABLE:phot}.

     The UVOT is a photon-counting device with a frame read-out time
of 11.0329 ms.  It is only able to record one photon per detector cell
during each read out. This results in coincidence losses at high count
rates.  For very high count rates, corresponding to $V \la 13.5$,
these losses are significant and can dramatically affect the
photometry, so coincidence loss corrections must be made.  We have
corrected all of our data for coincidence loss, however the afterglow
has $V > 19$, so coincidence losses are negligible, typically less
than 0.01 mag, for the afterglow.  Coincidence loss corrections,
however, are significant for the stars used to compute the aperture
corrections.

     The zero points used to transform the instrumental UVOT
magnitudes to Vega magnitudes were taken from in-orbit measurements as
obtained from the HEASARC {\sl Swift\/}/UVOT Calibration Database
(CalDB)\footnote{\url{http://swift.gsfc.nasa.gov/docs/heasarc/caldb/swift/}}
dated 2005-11-18 in the file {\sc swuphot20041120v102.fits}.  Colour
terms were not applied to the photometric calibrations, but
preliminary calibrations of on-orbit data suggest that they are
negligible.


\subsection{Ground-Based Data\label{SECTION:ground}}

\subsubsection{DFOSC\label{SECTION:dfosc}}

     The XRF~050416A optical afterglow was observed with the 1.54m
Danish telescope equipped with DFOSC\@.  DFOSC is focal reducer camera
based on a backside illuminated EEV/MAT CCD, providing a pixel scale
of $0\farcs39$ per pixel and a field of view of $13\farcm6 \times
13\farcm6$.  The DFOSC observations were carried out on five
consecutive nights starting on 2005 Apr 17.1 UTC (see
Table~\ref{TABLE:phot} for further details).  Data reduction was
performed following standard procedures (bias level subtraction and
sky flat fielding) running under IRAF\footnote{IRAF is distributed by
the National Optical Astronomy Observatories, which are operated by
the Association of Universities with National Science Foundation.}.

     Photometry was performed on the combined DFOSC images using the
{\sc SExtractor} software \citep{BA1996} with its default parameters.
Magnitudes and magnitude errors were determined using the MAG\_AUTO
option.  The DFOSC data were calibrated using the $BV\!{R_C}{I_C}$
field photometry of \citet{H2005}.  All of the stars that were in
common between the \citet{H2005} photometry and our {\sc SExtractor}
photometry were matched and used to compute photometric zero points
for each DFOSC image.  Colour terms were not used as they did not
improve the quality of the calibration.

\subsubsection{NOTCam\label{SECTION:notcam}}

     Near-infrared observations of XRF~050416A were taken on 2005 Apr
17 and 2005 May 22 with NOTCam mounted on the Nordic Optical
Telescope.  NOTCam is a multi-mode instrument that is based on a $1024
\times 1024$ HgCdTe ``HAWAII'' detector providing a scale of
$0\farcs233$ per pixel and a field of view of $4^\prime \times
4^\prime$.  Data reduction was done by following a standard procedure
for near-infrared imaging (shift and expose, subtract a scaled median
sky, divide by twilight flat, align and combine by adaptive sigma
clipping).  The individual images were aligned from the centroid of
the star 12335122+2104142 in the 2MASS catalogue \citep{SCS2006}.

     Photometry was performed on the NOT images using {\sc SExtractor}
with the default parameters in the same way as for the DFOSC images.
The NOTCam photometry was calibrated using star 12335122+2104142.  The
2MASS catalogue magnitude of this star is $K_s = 11.383 \pm 0.018$
\citep{SCS2006}.  Our infrared photometry is listed in
Table~\ref{TABLE:phot}.


\section{Results\label{SECTION:results}}

\subsection{The Decay Rate\label{SECTION:decay}}

     We use our photometry to constrain the rate of a power law decay
of the afterglow in each filter where there is multi-epoch photometry.
The $X$-ray data \citep{MPC2006} shows a change in the $X$-ray light
curve at 1450 seconds which \citet{MPC2006} interpret as the end of
energy injection.  In order to directly compare the
ultraviolet/optical/infrared decay rate to the $X$-ray decay rate we
only consider photometry with mid-point times more than 1450 s after
the BAT trigger.  This is the period that \citet{MPC2006} call Phase C
and corresponds to the time after the second $X$-ray break.

     We find $\alpha_B = +1.18 \pm 0.68$, $\alpha_V = -0.96 \pm 0.53$,
$\alpha_R = -0.57 \pm 0.14$, and $\alpha_I = -1.04 \pm 0.66$.  It is
not clear if the increase in the $B$-band flux at 144\,815 s ($\approx
1.7$ days) is real or a noise spike causing the source to appear
brighter than it really is.  The mean decay rate, excluding the
$B$-band data, is $\overline{\alpha} = -0.86 \pm 0.15$ (standard
error).  This is in good agreement with the $X$-ray decay rate of
$\alpha_X = -0.88 \pm 0.02$ at the same time.  Further, the decay
rates in each filter, excluding $B$, are all within 2$\sigma$ of the
$X$-ray value.  This suggests that, for $t > 1450$ s, there is no
spectral break between the $X$-ray and optical bands.  As a further
test we used the spectral energy distribution from
Section~\ref{SECTION:sed} to convert all the photometry to the $R_C$
band and fit a single power law.  This yielded a slope of $-0.75 \pm
0.19$, in good agreement with the mean slope that we determined above.
We prefer to adopt the mean slope since it does not depend on
uncertainties, and possible temporal variations, in the spectral
shape.

     The early time UVOT $V$-band photometry is consistent with the
afterglow having a constant $V$-band magnitude between 126 and 1868
seconds after the BAT trigger.  The decay rate in this interval is
$\alpha_V = -0.11 \pm 0.17$.  This is consistent with the $X$-ray
slope during Phase B ($-0.44 \pm 0.13$) \citep{MPC2006} at the
1.6$\sigma$ level.

\subsection{The Spectral Energy Distribution\label{SECTION:sed}}

\begin{figure}
\plotone{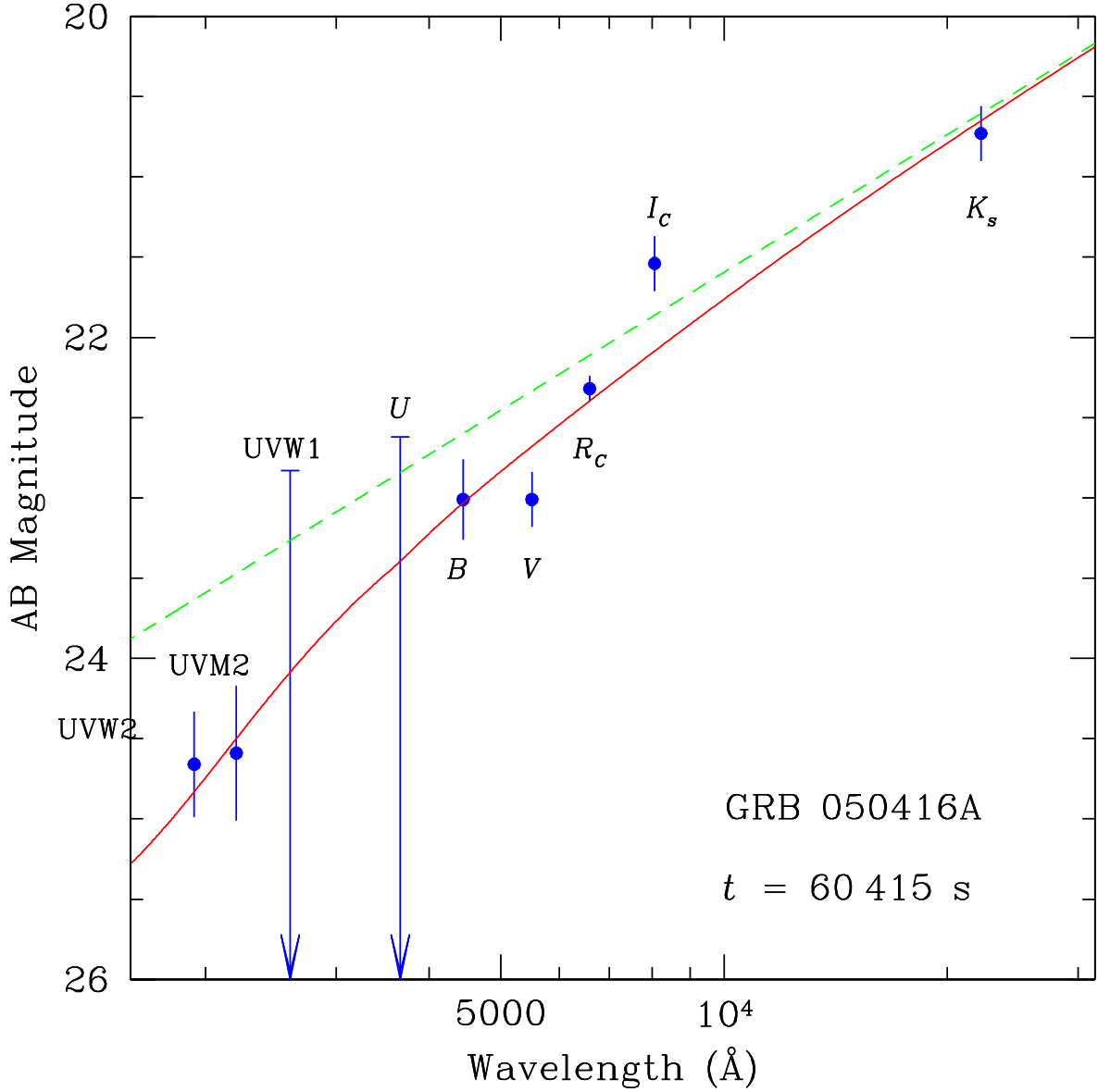} \figcaption[./f5.ps]{This Figure shows the SED of
the ultraviolet/optical/infrared afterglow of XRF~050416A at 60\,415 s
(0.7 days) after the BAT trigger.  The filled circles represent
observed photometry corrected for extinction in the Milky Way.  The
solid line represents the SED fit with an extinction in the host of
$A_V = 0.19 \pm 0.11$ mag assuming an intrinsic power law spectrum
with a slope of $\beta = -1.14 \pm 0.20$.  The dashed line shows the
unreddened spectrum.\label{FIGURE:sed}}
\end{figure}

     We constructed the spectral energy distribution between 1930
{\AA} and 22\,200 {\AA} at 60\,415 s after the BAT trigger.  The
photometry data nearest this time for each filter were transformed to
60\,415 s assuming a decay rate of $\alpha = -0.88$.  We adopted the
$X$-ray decay rate instead of the optical decay rate that we
determined in Section~\ref{SECTION:decay} because the $X$-ray decay is
better constrained, and we believe that the $X$-ray and optical decays
are the same (see Sect.~\ref{SECTION:decay}).  The UVW2 and UVM2
magnitudes were transformed to flux densities using the conversion
factors in the {\sl Swift\/}/UVOT CalDB\@.  The optical and infrared
magnitudes were converted to flux densities using the zero points of
\citet{FSI1995} and \citet{C2000}.  Each data point was corrected for
Galactic extinction using the reddening value of \citet{SFD1998} (see
Section~\ref{SECTION:obs}) but not for any extinction that may be
present in the host galaxy or in intergalactic space along the line of
sight to the burst.  The SED is shown in Figure~\ref{FIGURE:sed}.

     The SED was fit by $f_\nu(\nu) \propto \nu^\beta 10^{-0.4A(\nu)}$
where $f_\nu(\nu)$ is the flux density at frequency $\nu$, $\beta$ is
the intrinsic spectral index, and $A(\nu)$ is the extragalactic
extinction in the host ($z = 0.6535$) at frequency $\nu$. We found
that the best fit occurred for the Small Magellanic Cloud (SMC)
extinction law of \citet{P1992}, although using a Milky Way or Large
Magellanic Cloud extinction law does not significantly change the
results.  We prefer the SMC since it provides good fits to the
extinction seen in the host galaxies of other GRBs
\citep[\eg,][]{HWF2003}.  If we fix the intrinsic spectral slope to
the $X$-ray value of $\beta_X = -1.04$ \citep{MPC2006} the extinction
in the host is $A_V = 0.24 \pm 0.06$ mag.  If we allow both $\beta$
and $A_V$ to be free parameters the best fit occurs for $\beta = -1.14
\pm 0.20$ and $A_V = 0.19 \pm 0.11$.  Therefore, we believe that there
is approximately 0.2 mag of $V$-band extinction in the host along the
line of sight to XRF~050416A.

     The low value of the extinction in the host is at odds with the
high neutral hydrogen column density found by fitting the $X$-ray
spectrum \citep{MPC2006}.  \citet{PS1995} find $N_\mathrm{H} = (1.79
\times 10^{21}) A_V$ for the the conversion between hydrogen column
density and extinction in the Milky Way.  Using the value of
$N_\mathrm{H}$ derived from the {\sl Swift\/}/XRT $X$-ray observations
\citep{MPC2006} we find $A_V = 3.8^{+0.6}_{-0.7}$, which implies
significant extinction in the host along the line of sight to
XRF~050416A.  The discrepancy between this and the small amount of
extinction that we find from the ultraviolet/optical/infrared SED
suggests that the relationship between hydrogen gas and dust is
different in the host of XRF~050416A than it is in the Milky Way.
Combining the $X$-ray hydrogen column density with our estimate of
$A_V$ gives $N_\mathrm{H} = (36^{+21}_{-22} \times 10^{21}) A_V$ in
the host galaxy.  This is consistent with the $N_\mathrm{H}/A_V$
relation found in the SMC of $N_\mathrm{H} = (15.4 \times 10^{21})
A_V$ using Eq.~4 and Table~2 of \citet{P1992}.

     The high gas-to-dust ratio seen in the host of XRF~050416A is
typical of other GRB host galaxies, such as GRB~000301C
\citep{JFG2001}, GRB~000926 \citep{FGD2001}, and GRB~020124
\citep{HMG2003}, all of which have ratios consistent with that
observed in the SMC, \citep[see also][]{SFA2004,KKZ2006}.  This
suggests that our choice of using an SMC extinction law to determine
the intrinsic spectral slope is reasonable.  The high gas-to-dust
ratio may be a consequence of dust destruction by the ultraviolet and
$X$-ray flux from the GRB as described by
\citep[{\eg}][]{WD2000,PLF2003}.  Alternately, it may indicate that
the star formation in the host is fairly recent and that there may not
have been time for large amounts of dust to form.

\subsection{Cooling Break\label{SECTION:cooling}}

     The $V\!{R_C}{I_C}$ photometry yields a mean decay index of
$\alpha = -0.86 \pm 0.15$ for times more than 1450 seconds after the
burst.  The ultraviolet/optical/infrared SED at 60\,415 s has a slope
of $\beta = -1.14 \pm 0.20$, after correcting for extinction in the
host galaxy.  This is consistent with the intrinsic spectral slope
derived from the $X$-ray data.  The lack of evidence for a change in
either the intrinsic spectral slope or the decay rate between the
$X$-ray and optical bands suggests that there is no spectral break
between these two regimes.  Therefore the synchrotron cooling
frequency must lie either above the $X$-ray band or below the $K_s$
band at 60\,415 s ($\approx 0.7$ days) after the BAT trigger.

     The spectral and temporal decay rates can be used to predict the
index of the electron energy distribution, $p$.  The decay rate will
be a more robust number than the spectral index since the spectral
slope may be affected by uncertainties in the absorption along the
line of sight.  After the jet break the magnitude of the decay index
should equal the value of the electron index.  However, electron
indices of less than unity are unphysical.  Therefore, we can rule out
a jet break before 42 days after the BAT trigger because the $X$-ray
decay rate remains constant at $-0.88$ between 1450 s and at least 42
days.

     In order to determine the location of the cooling break, and to
estimate the value of the electron index, we use closure relations
between $\alpha$ and $\beta$.  These are based on the relations of
\citet{DC2001} when $1 < p < 2$.  For $p \ge 2$ the relations of
\citet{SPH1999} were used for the case of a homogeneous circumburst
environment, and the relations of \citet{CL1999} were used for the
case of a pre-existing stellar wind.  Table~\ref{TABLE:predict} lists
the closure relations and their observed values.  None of the closure
relations for $p \ge 2$ give values consistent with zero.  Therefore,
we believe that $1 < p < 2$.  This is a low value for a GRB, but three
of the ten bursts studied by \citet{PK2002} had $p \approx 1.4$, so it
can not be ruled out.  For $1 < p < 2$ the case of a homogeneous
circumburst medium with $\nu_X < \nu_c$, and the cases of a
pre-existing stellar wind with $\nu_X < \nu_c$ or $\nu_c <
\nu_\mathrm{opt}$ all give comparable closure values that are
consistent with zero.

\begin{deluxetable}{cclcll}
\tabletypesize{\scriptsize}
\tablewidth{0pt}
\tablecaption{This Table lists the closure relationships for the
various cases under consideration.  The theoretical value of each
closure relation is zero.\label{TABLE:predict}}
\tablehead{%
        \colhead{Case} &
        \colhead{Model} &
        \colhead{Environment} &
	\colhead{$p$} &
        \colhead{Closure} &
        \colhead{Value}}
\startdata
  1 & $\nu_{\mathrm{opt}} < \nu_X < \nu_c$ & ISM  & $1 < p \le 2$ & $\alpha - 3/8\beta + 9/16$ & $+0.13 \pm 0.19$ \\
  2 &                                      &      &  $p > 2$  & $\alpha - 3/2\beta$        & $+0.85 \pm 0.11$ \\
  3 &                                      & Wind & $1 < p \le 2$ & $\alpha - 1/4\beta + 9/8$  & $+0.55 \pm 0.16$ \\
  4 &                                      &      &  $p > 2$  & $\alpha - 3/2\beta + 1/2$  & $+0.35 \pm 0.11$ \\
  5 & $\nu_c < \nu_{\mathrm{opt}} < \nu_X$ & ISM  & $1 < p \le2$ & $\alpha - 3/8\beta + 5/8$  & $+0.19 \pm 0.19$ \\
  6 &                                      &      &  $p > 2$  & $\alpha - 3/2\beta - 1/2$  & $+0.55 \pm 0.11$ \\
  7 &                                      & Wind & $1 < p \le 2$ & $\alpha - 1/4\beta + 3/4$  & $+0.18 \pm 0.16$ \\
  8 &                                      &      &  $p > 2$  & $\alpha - 3/2\beta - 1/2$  & $+0.55 \pm 0.11$ \\
\enddata
\end{deluxetable}

     If $1 < p < 2$ then the predicted intrinsic spectral slope is
between 0 and $-0.5$ if the cooling break is above the $X$-ray band
and between $-0.5$ and $-1$ if the cooling break is below the optical
bands.  Both the $X$-ray and ultraviolet/optical/infrared spectral
slopes are consistent with $\beta = -1$, which corresponds to $p = 3$
for $\nu_X < \nu_c$, which is unusually high for a GRB.  However, if
$\nu_c < \nu_\mathrm{opt}$ then $p = 2$, which is fairly typical for
GRBs.  An electron index of $p = 2$ is consistent with cases 5 and 7
of the closure relations in Table~\ref{TABLE:predict}.  In these two
cases the predicted optical decay is $\alpha = -1$ regardless of the
density structure of the environment that the burst is expanding into.
This decay slope is slightly steeper than the observed decay rate of
$-0.86 \pm 0.15$, but it is within the observed uncertainties.
Therefore, we believe that $p \approx 2$ and $\nu_c <
\nu_\mathrm{opt}$ for XRF~050416A at 60\,415 s after the BAT trigger.

\subsection{Energy Considerations\label{SECTION:energy}}

     \citet{Sa2005} find that the isotropic equivalent energy of
XRF~050416A is $1.2 \times 10^{51}$ erg.  This, and the observed
limits on the jet break time, can be used to estimate the opening
angle of the jet and thus the total gamma-ray energy of the burst
\citep{R1999,SPH1999,FKS2001}.  The jet opening angle for XRF~050416A
is

\begin{equation}
  \theta_j = 0.106 t_j^{3/8} {(\eta_\gamma / 0.2)}^{1/8} {(n / 0.1)}^{1/8},
\end{equation}

\noindent
where $t_j$ is the observed jet break time in days since the BAT
trigger, $\eta_\gamma$ is the efficiency of converting energy in the
ejecta into gamma rays, and $n$ is the particle density in cm$^{-3}$.
The $X$-ray light curve \citep{MPC2006} suggests that there is no jet
break at times earlier than 42 days.  Setting $t_j \ge 42$ yields
$\theta_j \ge 25\arcdeg$ assuming $\eta_\gamma = 0.2$ and $n = 0.1$
cm$^{-3}$.  This result is not strongly sensitive to our choices of
$\eta_\gamma$ and $n$.  The corresponding energy in gamma rays,
corrected for beaming, is $E_\gamma \ge 1.1 \times 10^{50}$ erg.  The
upper limit on the gamma-ray energy can be obtained from the case
where there is no beaming (and thus no jet break).  In that case
$E_\gamma = E_\mathrm{iso}$, which implies that $1.1 \times 10^{50}$
erg $\le E_\gamma \le 1.2 \times 10^{51}$ erg.  Therefore the total
energy must be less than the canonical GRB energy of \citet{BFK2003}
if there is any beaming.

     \citet{Sa2005} showed that XRF~050416A fits onto the Amati
relation \citep{AFT2002} between $E_\mathrm{iso}$ and $E_p^\prime$.
They also showed that this burst can not be made to satisfy the
Ghirlanda relation \citep{GGL2004} between $E_\gamma$ and $E_p^\prime$
due to the lack of a jet break out to 35 days after the BAT trigger.
\citet{MPC2006} have shown that there is no jet break to at least 42
days, which makes the discrepancy with the Ghirlanda relation even
larger.

     The \citet{LZ2005} relation

\begin{equation}
  {E_\mathrm{iso} \over 10^{53}\,\mathrm{erg}} = 0.85
       \times {\left(E_p^\prime \over 100\,\mathrm{keV}\right)}^{1.94}
       \times {\left(t_j^\prime \over 1\,\mathrm{day}\right)}^{-1.24}
\end{equation}

\noindent
implies that the jet break should occur at $\approx 1.5$ days after
the BAT trigger in the observer's frame.  If we assume that
XRF~050416A has a two-component jet with a narrow component that
breaks at 1.5 days the observed decay between 1.5 and 42 days will be
the sum of the post-narrow-jet break component for the narrow jet and
the post-cooling break component for the wide jet.  A two-component
jet model has been proposed to explain some of the observed features
in the decays of GRB~030329 \citep{BKP2003,SFW2003}, such as the
presence of two achromatic breaks.  We propose that the observed
optical afterglow of XRF~050416A may be the sum of two jets that both
conform with the Ghirlanda relation and the Liang \& Zhang relation.
In this picture the narrow jet would have an opening angle of
$\theta_{j,n} \approx 7\arcdeg$ assuming $t_{j,n} = 1.5$ days, and the
wide jet would have $\theta_{j,w} \ga 25\arcdeg$ for a wide jet break
time of $t_{j,w} > 42$ days.  The gamma-ray energies in each jet are
$E_{\gamma,n} \approx 5 \times 10^{48}$ erg and $E_{\gamma,w} \ga 7
\times 10^{49}$ erg, which is roughly consistent with the division of
energy in the two-component jet model for GRB~030329.

     \citet{PKG2005} find that for two component jets where the total
energy in the wide jet is greater than that in the narrow jet the wide
jet will dominate the optical afterglow after approximately 0.1--1
days.  Since the deceleration time of the wide jet is similar to the
break time of the narrow jet the emergence of the wide jet at this
time can mask the steepening of the light curve caused by the jet
break in the narrow component.  We suggest that this is what has
happened in XRF~050416A.  We see no evidence for a change in the
spectral slope or the decay rate between the optical and $X$-ray
regimes at 60\,415 s ($\approx 0.7$ days).  This implies that both the
narrow and wide components have cooling frequencies below the optical
band at this time.

\subsection{Constraints on a Supernova Component\label{SECTION:supernova}}

     UVOT observation in the $V$-band filter were taken up to 74 days
after the BAT trigger.  We searched the late time exposure for
evidence of a rebrightening.  There is one low-significance UVOT
detection of the afterglow after 144\,815 s (1.676 d).  There is a
second low-significance detection at 3\,732\,868 s (43 d).  A visual
examination of the second detection suggests that it is a noise spike
in the data and not a point source.

     A type Ib/c supernova like SN1998bw \citep{PP1998} at the
distance of XRF~050416A is expected to peak at approximately $10(1+z)
\approx 16$ days after the burst.  We find no evidence for a source at
the location of the afterglow at this time down to a 3$\sigma$
limiting magnitude of $V_{\lim} = 21.9$ in coadded exposures.  For $z
= 0.6535$ the observed $V$ band approximately corresponds to the rest
frame $U$ band, so our upper limit corresponds to $M_U < -21.1$.
SN1998bw had a peak $U$-band absolute magnitude of $M_U = -19.16$
\citep{GVP1998} so the UVOT data is not able to constrain the
existence of a supernova component to the afterglow of XRF~050416A.


\section{Conclusions\label{SECTION:conc}}

     XRF~050416A is a near-by ($z = 0.6535$), short ($T_{90} = 2.4$ s)
XRF that shows no evidence for a jet break out to at least 42 days
after the burst.  The spectral slope and decay rate are the same in
the optical as in Phase C of the $X$-ray decay \citep{MPC2006}.  This
suggests that there is no cooling break between approximately 1.24
{\AA} (10 keV) and 22\,000 {\AA} (0.006 keV) at 60\,415 s after the
burst.  Further, the constancy of the decay suggests that the cooling
break did not pass through the optical between 1450 s (0.016 days) and
406\,438 s (4.7 days), nor did it pass through the $X$-ray regime
between 1450 s (0.016 days) and $\approx 3.6 \times 10^6$ s (42 days).
We find the best agreement with the synchrotron model occurs if the
cooling break is below the optical at 60\,415 s (0.7 days) after the
burst.

     The optical light decay slope is $\alpha = -0.86 \pm 0.15$ and
the intrinsic ultraviolet/optical/infrared spectral slope is $\beta =
-1.14 \pm 0.20$.  The best estimate of the electron index is $p
\approx 2$.  We are unable to distinguish between a burst occurring in
a homogeneous environment or a wind-stratified one.  The lack of a jet
break out to $\ga 42$ days implies that the jet opening angle is
$>25\arcdeg$ and the gamma-ray energy is $1.1 \times 10^{50}$ erg $\le
E_\gamma \le 1.2 \times 10^{51}$ erg.  If XRF~050416A has the
canonical energy of \citet{BFK2003} then the burst is not beamed.

     An alternate interpretation of the data is that XRF~050416A has a
two component jet.  The narrow component has an opening angle of
$\theta_{j,n} \approx 7\arcdeg$ and experienced a jet break at 1.5
days, in agreement with the prediction of the \citet{LZ2005} relation
between $E_\mathrm{iso}$, $E_p^\prime$, and $t_j^\prime$.  The wide
component does not break until at least 42 days after the burst and
must have an opening angle of $\ga 25\arcdeg$.  The wide component jet
contains approximately 14 times more energy than the narrow component
jet does.  This behaviour is similar to what was seen in GRB~030329
\citep{BKP2003,SFW2003}.

     We find no evidence for a large amount of extinction along the
line of sight to XRF~050416A.  The best fit
ultraviolet/optical/infrared spectrum suggests that $A_V \approx 0.2$
mag in the host galaxy.  This is inconsistent with the large hydrogen
column density implied by the $X$-ray spectrum.  If we assume that the
gas-to-dust ratio in the host is the same as that in the Milky Way the
implied extinction is $A_V = 3.8^{+0.6}_{-0.7}$ mag.  Using the
extinction derived from the optical data and the hydrogen column
density derived from the $X$-ray data the gas-to-dust ratio in the
host is $N_\mathrm{H}/A_V = 3.6^{+2.1}_{-2.2} \times 10^{22}$.  This
is dramatically larger than it is in the Milky Way, but consistent
with what is seen in the SMC, and in the host galaxies of other GRBs.


\acknowledgements

     The authors wish to thank Scott Barthelmy, and the GRB
Coordinates Network (GCN) for rapidly providing precise GRB positions
to the astronomical community.  This research has made use of the
NASA/IPAC Extragalactic Database (NED), which is operated by the Jet
Propulsion Laboratory, California Institute of Technology, under
contract with NASA\@.  This publication makes use of data products
from the Two Micron All Sky Survey, which is a joint project of the
University of Massachusetts and the Infrared Processing and Analysis
Center/California Institute of Technology, funded by the National
Aeronautics and Space Administration and the National Science
Foundation.  We acknowledge the use of public data from the Swift data
archive.  This paper is based, in part, on observations taken with the
Nordic Optical Telescope, operated on the island of Santa Miguel de la
Palma jointly by Denmark, Finland, Iceland, Norway, and Sweden in the
Spanish Observatorio del Roque de los Muchachos of the Instituto de
Astrof{\'\i}sica de Canarias.  STH would like to thank the DARK
Cosmology Centre for its hospitality while writing part of this paper.
The DARK Cosmology Centre is funded by the Danish National Research
Foundation.  This work is sponsored at Penn State University by NASA's
office of Space Science through contract NAS5-00136 and at MSSL by
funding from PPARC\@.  The research of JG is supported by the Spanish
Ministry of Science and Education through programmes
ESP2002-04124-C03-01 and AYA2004-01515.


  
\end{document}